# Structural and magnetic properties of Pr-alloyed MnBi nanostructures

P. Kharel[1,2], V. R. Shah[2], X. Z. Li[2], W.Y. Zhang[1,2], R. Skomski[1,2], J. E. Shield[2,3] and D. J. Sellmyer[1,2]

[1]Department of Physics and Astronomy, University of Nebraska, Lincoln, NE, 68588

[2]Nebraska Center for Materials and Nanoscience, University of Nebraska, Lincoln, NE, 68588

[3]Mechanical and Materials Engineering, University of Nebraska, Lincoln, NE 68588

**Abstract**

The structural and magnetic properties of Pr-alloyed MnBi (short MnBi-Pr) nanostructures with a range of Pr concentrations have been investigated. The nanostructures include thin films having Pr concentrations 0, 2, 3, 5 and 9 atomic percent and melt-spun ribbons having Pr concentrations 0, 2, 4 and 6 percent respectively. Addition of Pr into the MnBi lattice has produced a significant change in the magnetic properties of these nanostructures including an increase in coercivity and structural phase transition temperature, and a decrease in saturation magnetization and anisotropy energy. The highest value of coercivity measured in the films is 23 kOe and in the ribbons is 5.6 kOe. The observed magnetic properties are explained as the consequences of competing ferromagnetic and antiferromagnetic interactions.

**Introduction**

The intermetallic compound MnBi has long been a material of interest because of its unusual magnetic properties, attractive magneto-optical properties and complex phase diagram [1, 2]. The room-temperature ferromagnetic phase of MnBi is called the low temperature phase (LTP) which undergoes a coupled structural and magnetic (first-order) phase transition at 628 K and transforms into a paramagnetic phase, the high temperature phase (HTP)[3]. The LTP- MnBi shows large magnetocrystalline anisotropy at room temperature which exhibits an unusual temperature dependence when the temperature is lowered below room temperature. In contrast to most other ferromagnetic materials, the anisotropy energy (K) is negative at helium temperature, increases with increasing temperature, passes through zero at 100 K, and attains a maximum value of 22 M ergs/cm$^3$ at 490 K [1]. The high-temperature value of K is much higher than that of $Nd_2Fe_{14}B$, the most powerful (highest energy product) permanent magnet developed so far. In thin films, MnBi shows a large room-temperature magnetocrystalline anisotropy perpendicular to the film plane [4]. Other interesting properties of this material in thin-film form are the high value of transport spin polarization [5] and an unusual spin correlations leading to a Kondo effect when doped with heavy and noble metals such as Pt and Au [6,7]. These materials properties may be important for modern technological applications such as high-temperature permanent magnets, spintronics and high-density magnetic recording.

Although MnBi shows high anisotropy energy, the maximum energy product (7.1 MGOe at room temperature) reported so far is relatively low for most practical applications [8]. The main reason for this low energy product is that MnBi has a relatively low saturation magnetization $M_s$ (theoretical $M_s$ = 3.66 $\mu_B$/Mn ~ 712 emu/cm$^3$) as compared to other strong permanent magnet materials [6, 9]. The LTP- MnBi has the hexagonal NiAs structure with alternate Mn and Bi layers in ABAC stacking, where only the Mn atoms carry magnetic moments. There are also interstitial vacancy sites equivalent to Bi sites in the NiAs structure of MnBi unit cell. Half of the atoms in the MnBi lattice (Bi atoms) do not contribute to the observed ferromagnetic moment, although they contribute to the high value of magnetic anisotropy. These Bi atoms carry a small negative moment (-0.17$\mu_B$/Bi) reducing the net magnetization [6]. The magnetic structure in MnBi becomes even more complex when the large interstitial sites are occupied by Mn atoms, because the interstitial Mn moments couple antiparallel to the rest of the other Mn moments [6]. On the other hand, MnBi is highly susceptible to moisture and oxygen contamination and also undergoes a structural phase transition before the Curie temperature is reached. Therefore, extensive research efforts are underway to (i) increase the resistance to degradation (ii) increase the coercivity and magnetization and (iii) form a phase that is thermally stable and exhibits a high Curie temperature. These improvements might be achieved by alloying MnBi with a suitable third element. Here, we present our experimental investigation on how magnetic properties including magnetization, coercivity, anisotropy energy and Curie temperature of MnBi change when it is alloyed with a small amount of a light rare-earth element Pr. We have chosen Pr to investigate the sign of the couplings between the Mn 3d and Pr 4f moments [10].

**Experimental methods**

We have studied the effect of Pr alloying on the structural and magnetic properties of MnBi in two different types of structures, namely films and ribbons.

*MnBi-Pr Films*: We prepared MnBi and MnBi-Pr films with a range of Pr concentrations using e-beam evaporation and annealing. The films were deposited under high vacuum (base pressure $6\times10^{-9}$ Torr) on preheated glass (about 125 $^{o}$C) substrates. A multilayer sample was prepared by depositing Bi as a base layer over which alternate layers of Mn and Pr were deposited. The c-axis oriented Bi base-layer acts as a template to developing c-axis texture in the MnBi-Pr films. The rate of evaporation and the layer thicknesses were monitored by a quartz-crystal thickness monitor. The final thicknesses of the films were determined using x-ray reflectivity measurement. In order to allow sufficient time to diffuse Mn and Pr into the Bi template, the multilayer samples were kept at 290 $^{o}$C (just above the melting point 272 $^{o}$C of Bi) for two hours and then annealed at 410 $^{o}$C for one hour before being slowly cooled to room temperature. The elemental compositions of the samples were determined from the layer thicknesses which were later confirmed by energy dispersive x-ray (EDX) spectroscopy. The data presented here were taken on five thin film samples having elemental compositions $Mn_{55}Bi_{45}$, $Mn_{54}Pr_2Bi_{44}$, $Mn_{53}Pr_3Bi_{44}$, $Mn_{52}Pr_5Bi_{43}$ and $Mn_{50}Pr_9Bi_{41}$.

*MnBi-Pr Ribbons*: The MnBi and MnBi-Pr ribbons were prepared in a two-step process followed by an annealing. The first step was to prepare MnBi and MnBi-Pr ingots, and the second step was a rapid quenching of a melt of the ingots in a melt spinner. The MnBi and MnBi-Pr ingots were prepared by arc-melting where an appropriate amount of Mn, Bi and Pr metals were melted on a water-cooled Cu hearth. The ingots, thus prepared, were induction melted in a quartz tube and the melt was ejected onto the surface of a rotating copper wheel where it rapidly solidified into ribbons. The tangential speed of the rotating wheel was kept at 40 m/s for all the samples and the melt-spinning process was carried out in a chamber filled with a suitable pressure of high purity argon gas. In order to obtain the intended crystal structure, the ribbons were annealed in a tubular vacuum furnace (~$10^{-7}$ Torr base pressure) at 350 $^{o}$C for 4 hours. We also prepared powder samples by mechanically grinding the annealed ribbons in acetone. The powder samples were mixed with epoxy resin and then aligned in a magnetic field of about 90 kOe. The intended elemental compositions were estimated from the weights of the starting metal pieces. The ribbon samples presented here have elemental compositions $Mn_{52}Bi_{48}$, $Mn_{50}Pr_2Bi_{48}$, $Mn_{51}Pr_4Bi_{45}$ and $Mn_{48}Pr_6Bi_{46}$ as determined by EDX spectroscopy.

The structural properties of the samples were studied with x-ray powder diffraction in Rigaku x-ray diffractometer and microstructures of the samples were investigated using a Tecnai Osiris Transmission Electron Microscope (TEM). A Quantum Design SQUID magnetometer and a Physical Properties Measurement System (PPMS) were used to investigate the magnetic properties of the samples.

**Results and discussion**

Figure 1(a) shows the room-temperature x-ray diffraction patterns of the MnBi and MnBi-Pr films with Pr concentrations 2, 3, 5 and 9 atomic percent. All the patterns are consistent with the standard x-ray pattern of the hexagonal NiAs crystal structure. The presence of only (00*l*) peaks indicates that the films are highly c-axis textured. We note that the c-axis in MnBi is the easy direction of magnetization and is perpendicular to the film plane. Although there are weak peaks from unreacted elemental impurities, no secondary alloy phases were found within the limit of x-ray diffraction. In contrast, the x-ray diffractograms of the annealed Pr alloyed MnBi ribbons indicate that the samples contain LTP-MnBi with some traces of elemental and compound impurities. The observed diffraction patterns suggest that the ribbons are crystallized with randomly oriented crystallites. However, a close examination of the relative intensities of the Bragg reflections indicates a presence of weak *c*-axis texture, Fig. 1(b). This fact is supported by the x-ray patterns of the mechanically crushed powders, where the intensities of the (00*l*) peaks are much smaller than the intensities of the corresponding peaks in the ribbons, Fig. 1(c).

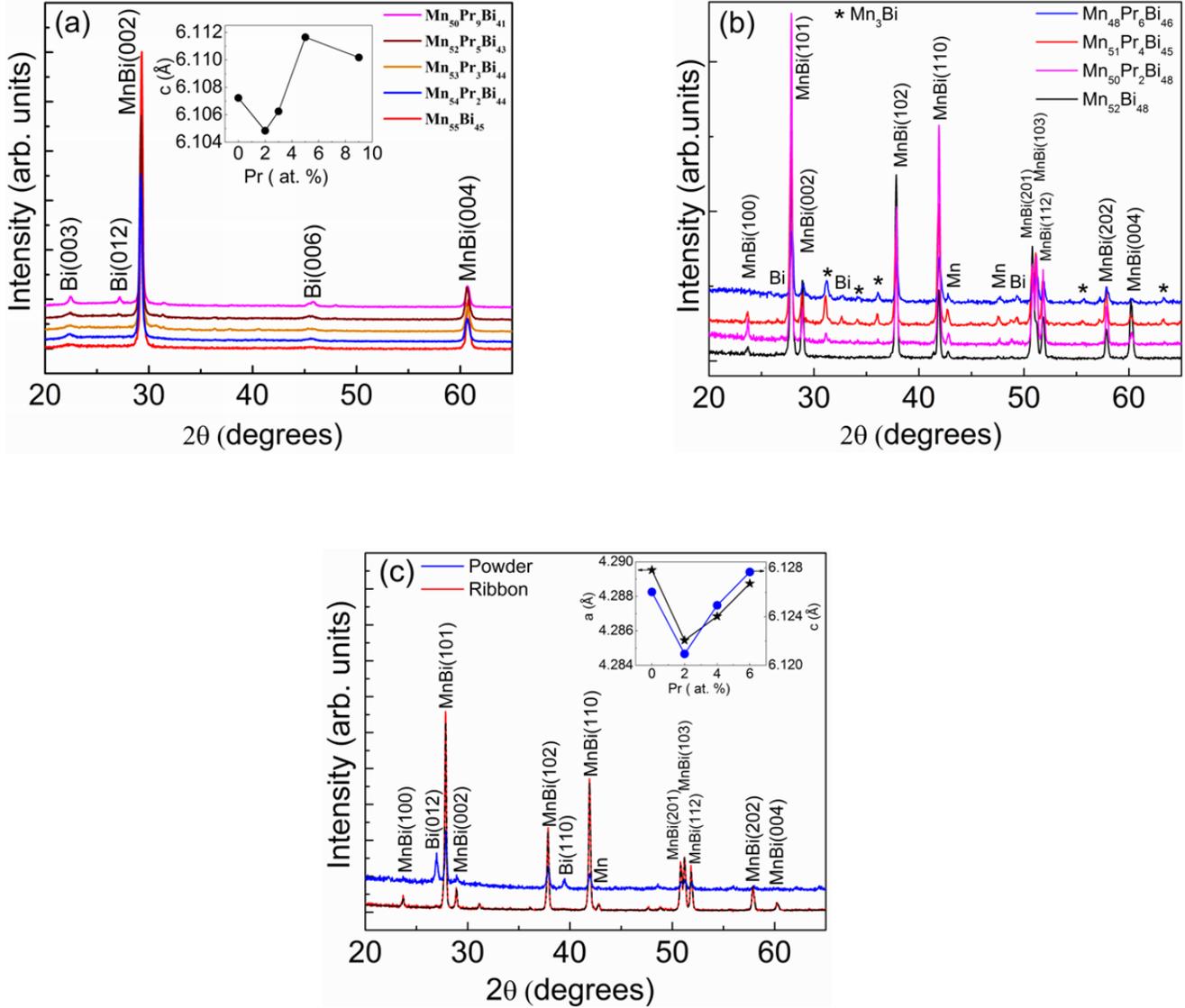

Fig. 1: Room-temperature x-ray diffraction patterns of (a) $Mn_{55}Bi_{45}$, $Mn_{54}Pr_2Bi_{44}$, $Mn_{53}Pr_3Bi_{44}$, $Mn_{52}Pr_5Bi_{43}$ and $Mn_{50}Pr_9Bi_{41}$ films and (b) $Mn_{52}Bi_{48}$, $Mn_{50}Pr_2Bi_{48}$, $Mn_{51}Pr_4Bi_{45}$ and $Mn_{48}Pr_6Bi_{46}$ melt-spun ribbons. Figure (c) shows the x-ray diffraction patterns of the ribbon and mechanically crushed powder samples of $Mn_{50}Pr_2Bi_{48}$ alloy. The lattice parameters as a function of Pr concentration for the thin film and the ribbon samples are shown as the insets

of figures (a) and (c) respectively. Figure (c) also shows a simulated powder x-ray diffraction pattern (Rietveld plot) corresponding to the hexagonal NiAs structure of the $Mn_{50}Pr_2Bi_{48}$ alloy as the dashed lines.

In order to understand the detailed crystal structure and to quantify the impurity phases, a Rietveld refinement of diffractograms of all ribbon samples was carried out using TOPAS software [11]. For the thin-film samples, the $c$-axis lattice parameters were determined from the analysis of the x-ray diffraction data considering Bi peaks as the reference. The refinement of the x-ray diffraction patterns of the ribbon samples shows that the samples mostly (more than 95 wt.%) contain LTP-MnBi with the traces of other elemental and alloy impurities. As shown in Fig. 1(b), the main impurities in the $Mn_{52}Bi_{48}$ ribbons are the elemental Mn and Bi but the samples alloyed with Pr also show the presence of a $Mn_3Bi$ secondary phase. We have included the crystallographic texture in the refinement model; however, the preferential occupation of Pr at the lattice sites could not be precisely determined because of the strong influence of the texture in the intensities of the Bragg peaks. As shown in the insets of Fig.1 (a) and (c), there is a small change in the lattice parameters of both MnBi films and ribbons due to Pr addition. Although we could not precisely determine the site occupancy of Pr atoms in the MnBi lattice by the Rietveld refinement of x-ray diffraction data, we expect that the Pr atoms may occupy the Bi sites because of similar atomic radii of Bi and Pr atoms ($R_{Bi}$ = 1.70 Å and $R_{Pr}$ = 1.82 Å) [12]. As we mentioned above, MnBi lattice contains empty interstitial sites at positions equivalent to Bi sites which allow more space to adjust relatively larger Pr atoms in Bi sites. Therefore, we do not expect a large change in the lattice parameters due to an addition of a small amount of Pr.

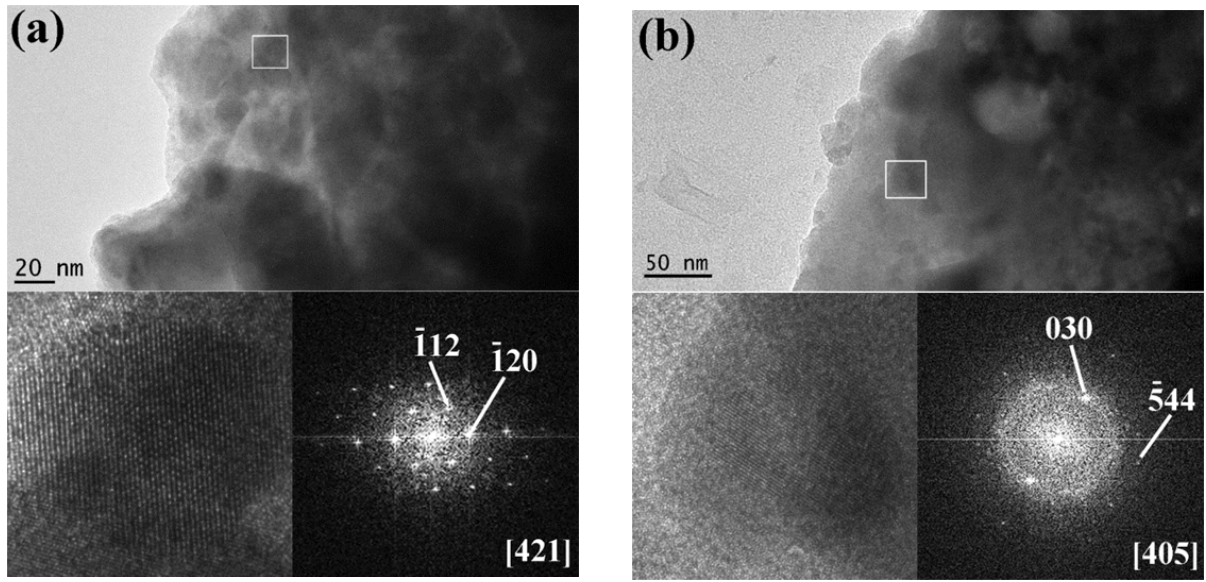

Fig. 2 Transmission electron microscope (TEM) images of the specimens prepared from (a) $Mn_{53}Pr_3Bi_{44}$ film and (b) $Mn_{50}Pr_2Bi_{48}$ ribbon. In both figures (a) and (b), the upper panels show the TEM bright field images, the lower left panels show the high resolution images and the lower right panels show the Fourier transform of the high resolution lattice fringes. The particles chosen for high-resolution image are marked by squares in the bright field images.

In order to better understand the microstructure and the effect of Pr on the crystal structure of MnBi, we have performed bright field and high-resolution TEM studies on two Pr alloyed samples: one from the ribbon group ($Mn_{50}Pr_2Bi_{48}$ ribbon) and the other from the film group ($Mn_{53}Pr_3Bi_{44}$ film). As shown in the upper panels of Figures 2(a) and (b), both the films and ribbons contain nanostructured particles. The size distribution of the particles in both the samples is not uniform and a lot of particles are overlapping. The average particle size in ribbon is about 20 nm and that in films is about 50 nm. These values are similar to the grain sizes estimated from the x-ray diffraction data using Scherrer's equation. However, the average grain size in $Mn_{53}Pr_3Bi_{44}$ films is about one half of the size measured in the MnBi films prepared under the similar conditions (see ref.13 for details). The high-resolution TEM images as shown in the lower-left panels of both figures (a) and (b) show that the particles are single grain or have overlapping grains. We did not find any evidence of Pr metal clusters in these samples. The crystal lattice parameters in both film and ribbon samples were confirmed using Fourier transform of the high resolution TEM images and found consistent with the values obtained from the x-ray diffraction data (see the lower-right panel of Fig. 2 for the Fourier transform).The zone axes and indexing of crystal planes were determined by comparing the patterns obtained from the Fourier transform of the electron diffraction pattern obtained from a simulation using SAEDZ software [14].

We have found a significant change in the magnetic properties of both the MnBi films and ribbons due to Pr addition. Figure 3(a) shows the out-of- plane M(H) hysteresis loops for the pure and Pr-alloyed MnBi films measured at room temperature. Although all the films show almost rectangular M(H) loops, the saturation magnetization and coercivity change considerably as Pr concentration increases. As shown in the inset of Fig. 3(a), the saturation magnetization decreases almost linearly from 569 emu/cm$^3$ to 169 emu/cm$^3$ as Pr concentration increases from 0 to 9%. But the coercivity shows an opposite trend and increases by a factor of about 9 from 2.6 kOe to 23 kOe for the same change in Pr concentration. On the other hand, the M(H) hysteresis loops of MnBi and MnBi-Pr ribbons are not completely saturated even at 70 kOe [Fig. 3(b)]. The high-field magnetization (at H = 70 kOe) of the pure MnBi (Mn$_{52}$Bi$_{48}$) ribbon is slightly higher (about 650 emu/cm$^3$) than that of the pure MnBi (Mn$_{55}$Bi$_{45}$) film, but the coerciviy is much smaller (0.5 kOe). However, there is a big increase in the coercivity of the MnBi ribbon due to Pr addition, where H$_c$ = 5.4 kOe for ribbons with 6% Pr concentration [Fig. 3(b) inset]. The change in the magnetization of the MnBi ribbon due to Pr addition is similar to that of the films. The high-field magnetization of the ribbon (M$_{7T}$) decreases by about 40% as the Pr concentration increases from 0% to 6%. This decrease in magnetization is comparable to the decrease observed in the films (50% decrease for 6% Pr concentration) [Fig. 3(a) inset]. We note that the magnetization in MnBi is sensitive to elemental composition (73 emu/g for $Mn_{52}Bi_{48}$ and 72 emu/g for $Mn_{50}Bi_{50}$ ribbons), but the big decrease in magnetization in the Pr-alloyed samples cannot be attributed to the small change in Mn to Bi ratio alone. In MnBi, it is very difficult to control the diffusion of Mn into the interstitial sites. As mentioned above, the interstitial Mn moments couple antiparallel to the rest of the MnBi lattice reducing the net magnetization [6]. In the present case, the external impurity atoms are expected to occupy the Bi sites with some of the Mn atoms in the interstitial sites. This leads to a magnetic structure in the Pr-alloyed MnBi compounds in which the Pr moments are only a small perturbation as compared with the major magnetization decrease resulting from the Mn on interstitial sites. This situation is similar to our previous theoretical and experimental investigations on Pt- and Fe-doped MnBi films reported in the references 6 and 13. This also suggests that the coupling between Mn and Pr moments may be antiferromagnetic.

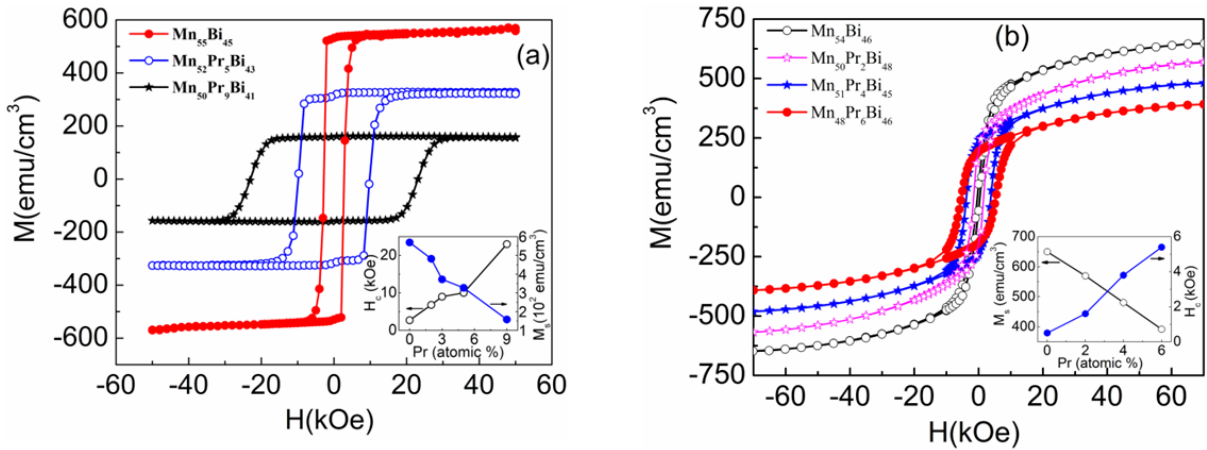

Fig. 3: (a) Room-temperature magnetization as a function of magnetic field of three selected MnBi-Pr films, where Pr content is 0, 5 and 9 percent respectively. Magnetic field was applied perpendicular to the film plane which is the easy direction of magnetization for MnBi. (b) Magnetic field dependence of room-temperature magnetization in MnBi-Pr melt-spun ribbons having Pr concentrations 0, 2, 4 and 6 percent respectively. Insets show coercivity ($H_c$) and saturation magnetization ($M_s$) versus Pr concentrations.

Since MnBi shows large perpendicular magnetic anisotropy in thin films, it is interesting to investigate the magnetic properties of MnBi particles after they are aligned in a magnetic field. As mentioned above, the ribbons were ground in acetone using ceramic mortar and pestle to crush into fine powder which was aligned in a magnetic field of 90 kOe. The fine powder contains particles with tens of microns in size. Figure 4 shows the room-temperature magnetization of $Mn_{51}Pr_4Bi_{45}$ particles as a function of magnetic field. This figure contains three M(H) loops; one for the loose particles and other two for the aligned particles. For the aligned particles, both the easy axis and hard axis magnetizations are shown. After mechanical grinding, an unexpected decrease in both the magnetization and coercivity was observed. The high field magnetization (367 emu/cm$^3$) of $Mn_{51}Pr_4Bi_{45}$ powder is about 23 % less than the value measured in the ribbon. The decrease in coercivity is even higher, from 3.8 kOe of the ribbon to 0.4 kOe of the powder. The decrease in magnetization can be attributed to the oxidation of the particles, presence of elemental Bi in the powder and also to the large demagnetizing factor of the particles. This is consistent with the x-ray diffraction result that the patterns for mechanically crushed powders contain intense peaks from elemental Bi, Fig. 1(c). It is likely that the decrease in coercivity of the mechanically ground powders is caused by the decrease in magnetic anisotropy and also by the destruction of the texture. We note that the anisotropy energy of the $Mn_{51}Pr_4Bi_{45}$ powder is about 40% less than that measured in the ribbons and the intensity of (00*l*) peaks in the x-ray diffraction pattern of the mechanically milled power is a lot smaller than that in the ribbon.

Although we observe clear anisotropy between the in-plane and out-of-plane loops, the fairly wide in-plane M(H) loop suggests that only a fraction of the particles are aligned with their c-axis perpendicular to the substrate plane, along which the external magnetic field was applied during the alignment. Since the ribbons were crushed into powder using mortar and pestle, the particles are not expected to be single domain. Therefore, we could not obtain perfect alignment of the particles with square M(H) loops as seen in the films. However, the alignment has produced a significant improvement in both the coercivity and squareness ratio ($M_r/M_s$). The out-of-plane coercivity as high as 8.7 kOe and $M_r/M_s$ ratio of about 80% have been measured in the aligned $Mn_{51}Pr_4Bi_{45}$ sample. These values are much larger than the values measured in the ribbon.

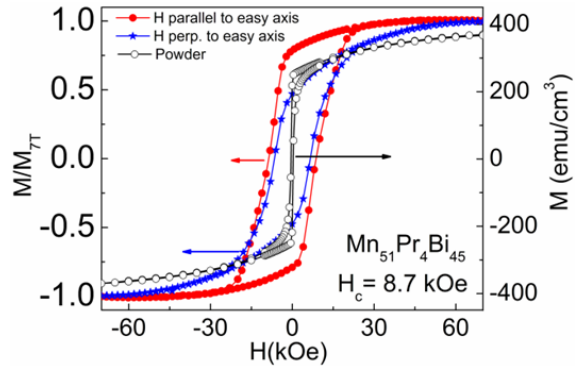

Fig. 4: Room-temperature M(H) hysteresis curve of $Mn_{51}Pr_4Bi_{45}$ powder prepared by mechanical grinding of melt-spun ribbons (scale on the right). M(H) curves of the $Mn_{51}Pr_4Bi_{45}$ powder mixed with epoxy-resin which was aligned in magnetic field are also shown in the same figure with the scale on the left.

As shown in the insets of Fig. 5(a) and 5(b), the anisotropy energy of MnBi decreases almost linearly with increasing Pr concentration, irrespective of the nature of the nanostructures. The anisotropy constant K was estimated using the approach to saturation method where the high-field data was fitted to $M = M_0(1-A/H^2) + \chi H$; $A = 4K^2/15M_0^2$. The parameters $M_0$, A and χ are the spontaneous magnetization, a constant that depends on K and the high-field susceptibility, respectively [15]. This suggests that the increase in coercivity in Pr doped samples cannot be explained in terms of the change in anisotropy energy K. Since coercivity is not an intrinsic magnetic property and depends also on the microstructure of a sample, it is difficult to determine the exact mechanism leading to coercivity in a ferromagnetic material [16]. However, an empirical relation of the type $H_c = \alpha\, 2K_1/M_s - N_{eff} M_s$ is commonly used to analyze the properties of a hysteresis loop and explain the origin of coercivity [17, 18]. In this equation $H_c$, α, $K_1$, $M_s$ and $N_{eff}$ are the coercivity, a microstructural parameter, anisotropy energy, local effective demagnetization factor and spontaneous magnetization, respectively.

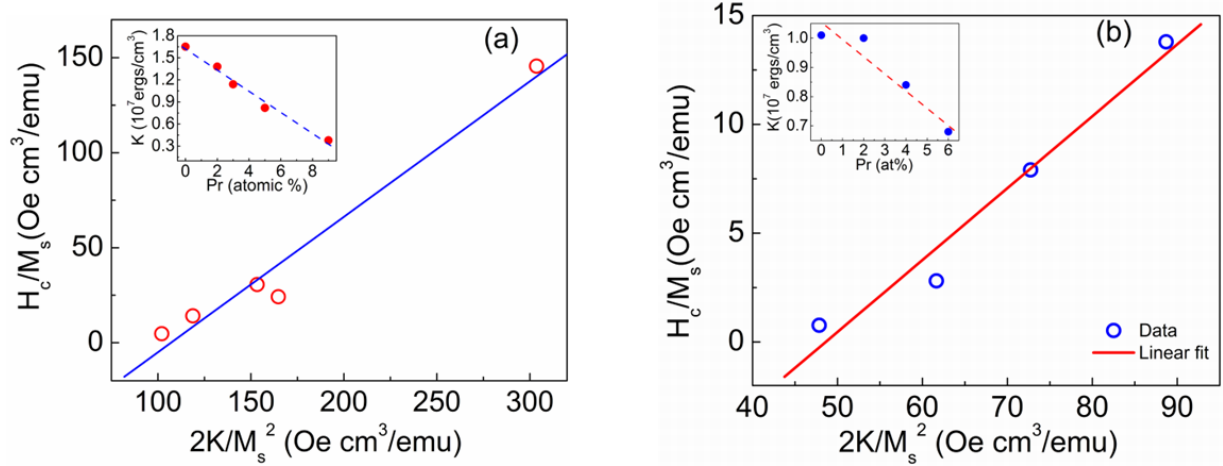

Fig. 5: Figures (a) and (b) plot $H_c/M_s$ versus $2K/M_s^2$. The data points in (a) were taken from the room-temperature M(H) curves of MnBi:Pr films having Pr concentrations 0, 2, 3, 5 and 9 percent respectively and that in (b) were taken from the M(H) curves of the ribbons having Pr concentrations 0, 2, 4, and 6 percent respectively. The inset plots the anisotropy constant K as a function of Pr concentration. The dashed lines are guides to the eye.

In general, $H_c/M_s$ ratios at various temperatures are plotted as a function of the corresponding values of $2K_1/M_s^2$ to quantitatively determine the value of α, which is related to the microstructure of the sample. Here, we have taken different approach and have plotted $H_c/M_s$ for different concentrations of Pr as a function of the corresponding values of $2K_1/M_s^2$ measured at room temperature. As shown in Fig. 5(a), the $H_c/M_s$ of MnBi-Pr film scales almost linearly with $2K_1/M_s^2$. The linear fit to the data points yields a value of α = 0.7. The ribbons also show similar behavior but with relatively smaller value of α, where α = 0.3. Although these values of α suggest that the leading mechanism of coercivity in the MnBi-Pr films is the nucleation of reverse domains and that in the ribbons is domain wall pinning [16], the big difference in the coercivities of these two types of nanostructures for the same elemental compositions cannot be explained in terms of these mechanisms alone. We believe that other factors such as c-axis texture of the films, particle size, and pinning of non-magnetic inclusions may play an important role in determining the observed large coercivity in the Pr-alloyed MnBi films. This is consistent with our x-ray diffraction and TEM studies as discussed above.

From the plots shown in Fig. 5, we also have determined the effective demagnetization factors. The values of $N_{eff}$ for the films and the ribbons are 24π and 4.5π respectively. Since the values of $N_{eff}$ depend on the shape, size and environment surrounding the grains, different values of $N_{eff}$ are expected for the films and ribbons but the observed values of $N_{eff}$ are too large to have a simple interpretation. Kronmüller *et al*. have also reported similar large values for the NdFeAlB magnets and have suggested that such large values of $N_{eff}$ are expected near nonmagnetic inclusions and sharp edges of the grains [18]. In our case, both the films and ribbons contain non-magnetic elemental Bi and this may contribute to contribute to the observed large values of $N_{eff}$. The relatively smaller value of $N_{eff}$ in the Pr alloyed-MnBi ribbons suggests that the amount of non-magnetic impurity, mainly Bi, is smaller in the ribbons as compared to that in the films. This is consistent with the XRD results and also with the fact that the saturation magnetization in the MnBi film is only 87% of the high field magnetization measured in the corresponding ribbons.

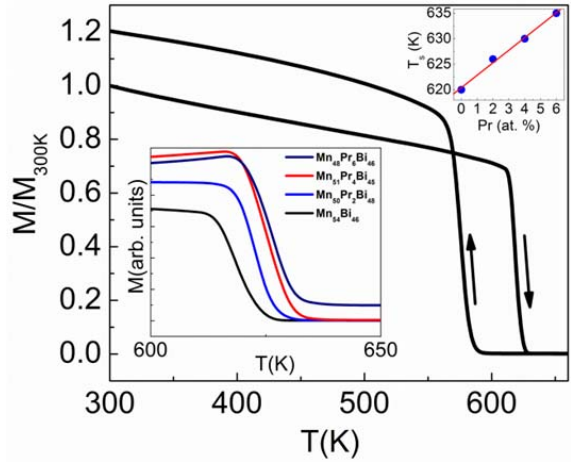

Fig. 6: Magnetization as a function of temperature M(T) of $Mn_{52}Bi_{48}$ ribbon measured at H = 1 kOe during heating from room temperature to 750 K and cooling from 750 K to room temperature. The inset at the bottom-left corner shows the M(T) curves of Pr-alloyed MnBi ribbons measured at 1 kOe during heating and the inset at the top-right corner shows structural phase transition temperature ($T_s$) as a function of Pr concentration.

In order to understand the effect of Pr on the structural phase transition temperatures of MnBi, we have measured the temperature dependence of magnetizations of MnBi and MnBi-Pr ribbons between 300 K and 750 K. As shown in Fig. 6, $Mn_{52}Bi_{48}$ ribbons undergo a coupled structural and magnetic phase transition from the ferromagnetic low-temperature phase (LTP) to the paramagnetic high-temperature phase (HTP) at 620 K when the sample is heated from room temperature. When the sample is cooled from 750 K, the sample transforms from HTP to LTP at 580 K showing a thermal hysteresis at the phase transition. Also, the magnetizations during cooling are little larger than the magnetizations during heating. Since the applied field of 1 kOe is not sufficient to saturate the magnetization below the Curie temperature, it is possible that there is a magnetic domain rearrangement as the samples pass through HTP to LTP at a magnetic field resulting in an increase in magnetization. The phase transition temperatures were determined from the extrapolation of the straight portions of the M versus 1/T curves. These values are very close to the transition temperatures (633 K for LTP to HTP and 603 K for HTP to LTP) determined from the neutron diffraction studies of bulk MnBi [19]. Although all the Pr-alloyed samples also show thermal hysteresis at the phase transition, the phase transition temperatures are different and are shifted to the higher values. As shown in the insets of Fig. 6, the LTP to HTP transition temperature ($T_s$) shows a linear increase from 620 K to 635 K as the concentration of Pr in MnBi increases from 0 to 6 at.%. However, the real Curie temperature of these alloys could not be determined as the samples underwent structural phase transitions before the Cure temperature was reached. The linear increase in $T_s$ with Pr concentration supports the fact that Pr may occupy MnBi lattice sites instead of segregating as secondary alloy phases or metal clusters.

**Conclusions**

We have studied the effect of Pr on the structural and magnetic properties of MnBi in two types of nanostructures namely films and ribbons. In both the cases, the saturation magnetization decreases and the coercivity increases due to Pr addition. X-ray diffraction shows that the film samples are highly c-axis textured but the ribbons are nearly polycrystalline with small c-axis texture. TEM investigation shows that the samples contain nanostructured single-grain particles. In both the films and ribbons Pr alloying produced a significant change in the magnetic properties of MnBi nanostructures including increase in coercivity and decrease in magnetization and anisotropy energy. Interestingly, Pr-alloying stabilized the low-temperature phase where the phase transition temperature increased from 620 K to 630 K as the Pr concentration increased from 0 to 6%. The decrease in magnetization in MnBi-Pr alloys is qualitatively similar to that observed when MnBi is alloyed with Pt, Au and Fe. This is explained as an increase of antiferromagnetic interaction between Mn atoms on interstitial sites and the original Mn lattice. The magnetic effect of Pr moments appears to be a small effect in comparison with the large decrease of magnetization from antiferromagnetic interaction of interstitial and regular lattice Mn moments. The increase in coercivity with the increase in Pr concentration in both the films and ribbons cannot be explained on the basis only of anisotropy change but rather is attributed to the structure and nonmagnetic inclusions in the samples. Since the Pr-alloyed films show very high coercivity (23 kOe with 6% Pr), this material may be useful for the nanocomposite magnets with MnBi-Pr as the hard component.

**Acknowledgements**

This research is supported by NSF MRSEC (NSF-DMR-0820521) (PK), ARPA-E (DE-AR0000046, Subward No. 22101) (DJS, JES, RS, WYZ), ARO (W911NF-10-2-0099) (VRS) and NCMN (Central Facility support).